\documentclass[manuscript,screen]{acmart}

\AtBeginDocument{%
  }

\setcopyright{acmcopyright}
\copyrightyear{2018}
\acmYear{2018}
\acmDOI{XXXXXXX.XXXXXXX}

\acmConference[Conference acronym 'XX]{Make sure to enter the correct
  conference title from your rights confirmation emai}{June 03--05,
  2018}{Woodstock, NY}
\acmPrice{15.00}
\acmISBN{978-1-4503-XXXX-X/18/06}

\usepackage{color, colortbl}
\definecolor{teagreen}{rgb}{0.9, 0.97, 0.90}
\definecolor{wildblueyonder}{rgb}{0.64, 0.68, 0.82}

\usepackage{caption}
\usepackage{subcaption}
\usepackage{color,soul}

\begin{document}

\title{Context-Aware Sequential Model for Multi-Behaviour Recommendation}

\author{Shereen Elsayed}
\email{elsayed@ismll.uni-hildesheim.de}
\affiliation{%
  \institution{Information Systems and Machine Learning Lab, University of Hildesheim}
  \country{Germany}
}

\author{Ahmed Rashed}
\email{ahmed.rashed@vwfs.io}
\affiliation{%
  \institution{Volkswagen Financial Services}
  \country{Germany}
}

\author{Lars Schmidt-Thieme}
\email{schmidt-thieme@ismll.uni-hildesheim.de}
\affiliation{%
  \institution{Information Systems and Machine Learning Lab, University of Hildesheim}
  \country{Germany}
}

\settopmatter{printacmref=false}
\renewcommand{\shortauthors}{Elsayed, et al.}


\begin{abstract}
  Sequential recommendation models have recently become a crucial component for next-item recommendation tasks in various online platforms due to their unrivaled ability to capture complex sequential patterns in historical user interactions. Nevertheless, many recent sequential models mainly focus on modeling a single behavior, representing the platform's target relation, e.g., purchase. While on the other hand, other implicit user interactions, such as click information, and add-to-favorite, can provide deeper insights into the users' sequential behavior and allows better modeling of the users' profiles. Recent work in multi-behavioral models has been trying to partially address this problem by focusing on utilizing graph-based approaches for modeling multi-behavior data as heterogeneous graphs. However, many fail or neglect to capture the sequential patterns simultaneously. While few recent time-aware multi-behavioral methods try to address both aspects at the same time, they still consider auxiliary behaviors of the same importance to the learning process, which might not be the case in many scenarios. In this work, we propose a \textbf{C}ontext-\textbf{A}ware \textbf{S}equential \textbf{M}odel (CASM) for multi-behavioral recommendations that leverages the advantages of sequential models and can support an arbitrary number of behaviors seamlessly. Specifically, context-aware multi-head self-attention layers are employed to capture the multi-behavior dependencies between the heterogeneous historical interactions. Furthermore, we utilize a weighted binary cross-entropy loss to weigh the different behaviors differently through the learning process of the model to allow more precise control of their contributions based on the target recommendation scenario. Experimental results on four real-world datasets show that the proposed model significantly outperforms multiple multi-behavioral and sequential recommendation state-of-the-art approaches.
\end{abstract}




\maketitle

\section{Introduction}
The importance of recommender systems has grown massively in the last decade; recommender systems have been categorized into several categories, which include but are not limited to sequential recommendation, multi-behavioral recommendation, fashion recommendation, attribute-aware recommendations depending on the input data availability, e.g., items attributes, users attributes, timestamp and the target task like next item recommendation, next basket recommendation, and click-through rate prediction.

Many efforts have been made to propose a better user experience through more accurate recommendations, ranging from early studies about matrix factorization methods \cite{koren2009matrix} and recent studies about context-aware sequential models \cite{rashed2022context} and multi-behavioral approaches\cite{xia2021knowledge,yuan2022multi,yang2022multi}.
Furthermore, deep learning approaches started to emerge to enhance collaborative filtering methods, e.g., NCF \cite{he2017neural}, which uses superficial, fully connected, non-linear layers to model the user-item interactions. Some of the later work concentrates on the sequential pattern in the historical user-item interactions, i.e., sequential recommendation, which is a crucial area that faces many challenges, e.g., 1) How to model dynamics in the input sequence?, 2) How to perform a personalized next item recommendation based on individual historical preferences? Recently a surge of models was presented for sequential recommendations, tackling these challenges in different manners. Earlier methods, such as Caser \cite{tang2018personalized}, rely on convolutional filters; other methods employ RNNs, such as GRU4Rec \cite{jannach2017recurrent} as they are popular in sequence modeling. Recent approaches utilize the transformer architecture, e.g., SASRec\cite{kang2018self}, one of the first methods to use a multi-head self-attention block for modeling the sequence of historical interactions. Similarly, BERT4Rec \cite{sun2019bert4rec} applies a bi-directional attention mechanism for modeling historical user-item interactions. Lastly, CARCA \cite{rashed2022context} employs a multi-head self-attention with cross-attention to capture these historical patterns.

Although these models have shown superior performances in several scenarios, they only focus on a singular type of user-item interaction. On the other hand, click data can represent not only the primary user-item interaction but also other user behaviors over time. However, most existing recommender models focus on modeling the target/main behavior between users and items. In the last few years, the focus started emerging on the existence of multiple relations, which provides additional insight into user-item interactions. There are numerous online e-commerce websites where users have few sales interactions but a higher count of other interactions like views, likes, and add-to-favorite interactions. Furthermore, in the automotive industry, particularly in B2B used-cars centers, dealers have different behavioral interactions with different frequencies, such as vehicle sales, bids, and views. Therefore, considering other behaviors can be specifically beneficial for users who might have few historical sales but have other interactions which can be used to generate more expressive latent representations of their behavior.

Recently proposed methods started to utilize the click data as auxiliary information about the user-item interaction \cite{jin2020multi,xia2020multiplex,xia2021knowledge,xia2021graph,yuan2022multi,yang2022multi,rashed2020multirec}. Early multi-behavioral graph-based approaches \cite{jin2020multi,xia2021graph} that did not incorporate the time aspect suffered from poor predictive performance compared to recent time-aware sequential models. However, the latest multi-behavioral models \cite{yang2022multi,yuan2022multi} started to include the time factor, considerably affecting the model performance compared to the earlier methods. In this work, we benefit from the sequential aspect of the data, we propose a context-aware sequential method for a multi-behavior recommendation which combines different components and weighting functions to build a simple, flexible, and efficient model that can scale to any arbitrary number of behaviors that might exist in a recommendation setting while having minimal impact on training and inference/response time which is a strict functional requirement in many current online applications. Our contributions can be summarized as follows:
\begin{itemize}
\item \sloppy We propose a simple context-aware sequential method (CASM) for a multi-behavior recommendation, which is capable of capturing the sequential patterns in multi-behavioral data by employing self-attention layers.

\item We employ a weighted binary cross-entropy loss for weighing behaviors differently, which allows the model to have a fine-controlled contribution weight for each behavior.

\item We conducted extensive experiments on four multi-behavioral datasets, which show that the proposed method significantly outperforms current state-of-the-art methods on the four datasets and improves the results by a margin that reaches up to 19\% for multi-behavioral recommendation tasks. 

\item We analyze the batch runtime for several state-of-the-art methods, and our model shows more than 30 times faster than the closest competitor (MBHT \cite{yang2022multi}).
\end{itemize}

\section{Related Work}
 Related work can be split into three categories, context-aware recommendation models, sequential recommendation, and multi-behavior recommendation models.

 \textbf{Context-Aware Models} is a substantial sector in recommender systems that includes the contextual features in the recommender model. Early approaches utilizing contextual features are the factorization machine-based methods such as FM \cite{rendle2010factorization}. Later the neural factorization machine (NFM) utilized deep neural networks for learning non-linear feature interactions \cite{he2017neuralNFM}. Furthermore, DeepFM \cite{guo2017deepfm} is a popular context-aware method that combines the FM and DNNs for extracting latent representations. More recent approaches benefit from incorporating the time-aspect, which allows learning the sequential pattern in the data along with the contextual features. A state-of-the-art attribute and context-aware sequential method is the CARCA model \cite{rashed2022context}, which includes contextual information and item attributes. The sequence of profile items and the target items passes through a cross-attention mechanism to get the final items' scores.

\textbf{Sequential Recommendation} is a prevalent task in recommender systems that utilizes the historical interactions of each user to predict the next item the user will most likely interact with. GRU4Rec \cite{jannach2017recurrent} is one of the early approaches that utilize RNNs for mining the sequential behavior in a session-based recommendation scenario. Another sequential model is Caser \cite{tang2018personalized}, which applies convolutional filters on the time and item embeddings to capture the latent sequential behavior of users. Additionally, more recent approaches started to employ transformer architectures such as the SASRec \cite{kang2018self} model, which feeds the item embedding sequence into a multi-head self-attention block to capture the sequential correlation between historical items. Another method that improves upon SASRec is BERT4Rec \cite{sun2019bert4rec}, which uses a Bi-directional self-attention mechanism to better model the sequential behaviors. Other recent approaches tried to extend the  SASRec model by either adding personalized latent user features in the SSE-PT \cite{zhou2020s3}, adding the ability to model the time intervals between the sequence of interactions in TiSASRec \cite{li2020time}, and adding the ability to handle categorical features in the $S^3$Rec \cite{wu2020sse} model. 

With the current massive data sources available for recommender systems, recent models started not only to employ the user-item primary interaction relation such as purchases but also to use other available click data. Employing auxiliary information has different namings in the existing prior work: relation-aware recommender systems, multi-relation recommendation, and multi-behavior recommendation. 

Many recent \textbf{multi-behavior recommendation}  approaches rely on convolutional graph networks by treating the different behaviors as a global heterogeneous graph, such as the MB-GCN \cite{jin2020multi} and MB-GMN \cite{xia2021graph} models. Additionally, other methods tried to employ transformer encoders in their architecture for better representation learning. One such example is the memory-augmented transformer networks (MATN) \cite{xia2020multiplex}, which uses a transformer encoder followed by a cross-behavior aggregation layer to model the different behaviors. Moreover, KHGT \cite{xia2021knowledge} is a graph transformer method that captures user-item interaction characteristics; also, a graph attention layer is employed to understand the item-item relation further. Furthermore, a recent model showing state-of-the-art results is MBHT \cite{yang2022multi}, which uses low-rank self-attention followed by a hyper-graph neural architecture to capture the long-short term dependencies. Finally, MB-STR \cite{yuan2022multi} employs a multi-behavior sequential transformer layer to simultaneously learn sequential patterns among several historical behaviors. 

To this end, we propose a simple multi-behavioral context-aware sequential model (CASM) that employs multi-head self-attention layers for modeling historical sequential behaviors and leverages behavioral information as a contextual feature for better modeling and representation learning of the behavioral dependencies. This also allows CASM to be employed in diverse multi-behavioral settings with any arbitrary number of behaviors. Finally, CASM is trained with a weighted cross-entropy loss function to allow a fine-tuned control of the contribution of each relation in the multi-behavioral scheme.

\section{Problem Formulation}
\makeatletter
\DeclareRobustCommand*\cal{\@fontswitch\relax\mathcal}
\makeatother

 In sequential multi-behaviour item recommendation task we define a set of users ${\cal U}:=\{1,\ldots,$ $U\}$ and items ${\cal V}:= \{1,\ldots,I\}$, where each user has a sequence of items $S^u=[ (v_1, c_1), (v_2, c_2), ..., (v_{ |S^u|}, \\c_{|S^u|})]$ which represents the chronologically ordered interaction sequence, where each item interaction $v_j$ is associated with item behavior type $c_j$ such as add-to-cart or buy behaviour. 

The main goal of the task will be to rank a target list of items based on their likelihood of being interacted with by the target user $u$ at time $t + 1$ using all of his historical interactions until time step $t$.

In this scenario, we consider the user-item primary behavior (relation) our target behavior, such as the buying behavior. On the other hand, the other behaviors act as auxiliary behaviors, which provide additional information for better user behavior modeling. 


\begin{figure*}[ht]
  \centering
  \includegraphics[scale=0.425]{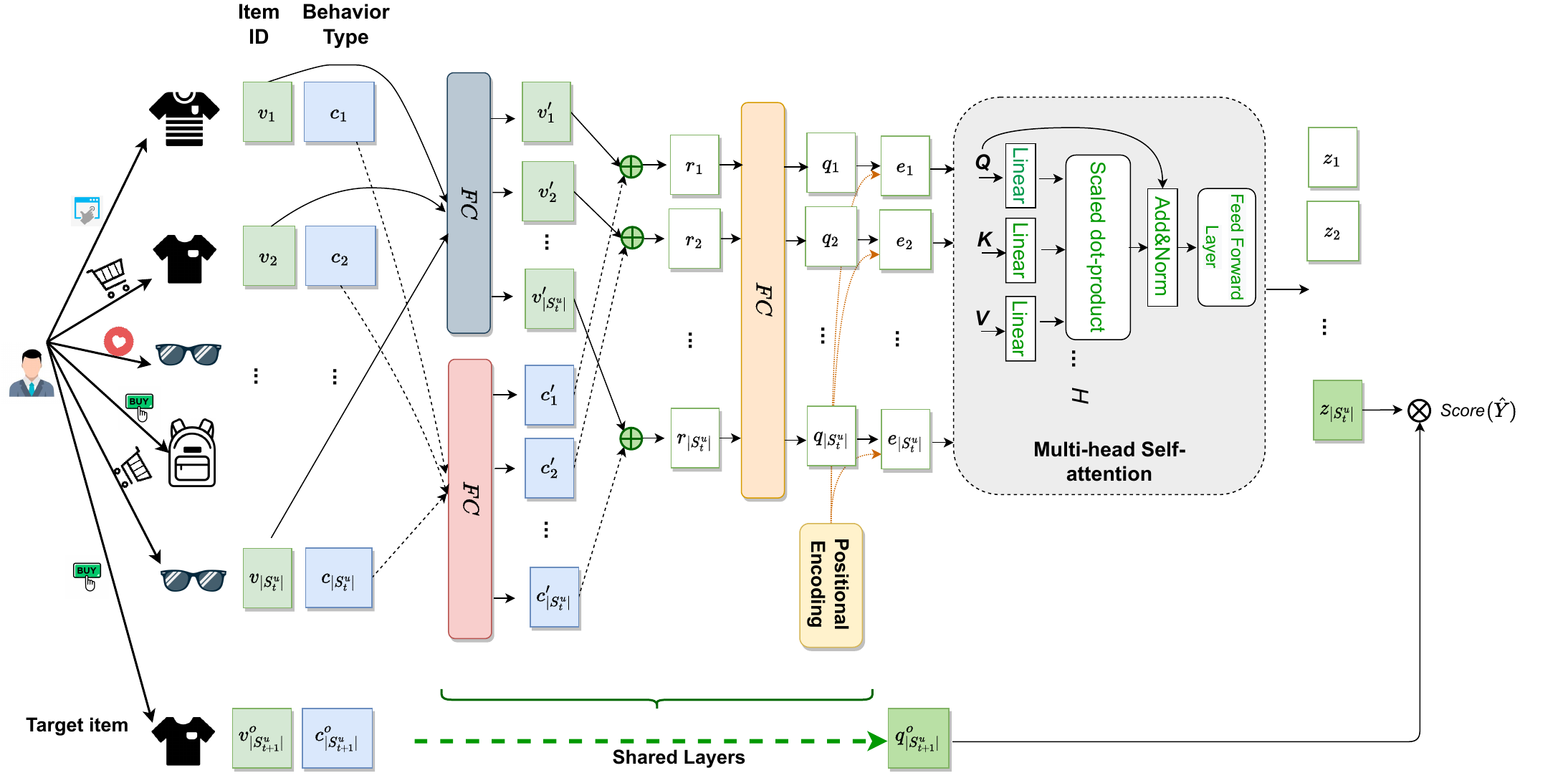}     
  \caption{Context-Aware Sequential Model Architecture}
  \label{fig:1}
\end{figure*}

\section{Methodology}

In this section, we elaborate on the different components of our proposed method, illustrated in Figure \ref{fig:1}. The model consists of three main components the item and context embedding layer, the multi-head self-attention, and the scoring components. 

\subsection{Items and Context Embedding}
In sequential recommendation settings, each user has a sequence of interactions over time, and we aim to recommend the next most likely item to be interacted with. For the multi-behavioral recommendation scenario, we consider not only the primary user behavior, like historical sales but also all other types of interactions that have occurred over time.
For each item in the sequence, there is a corresponding behavior representing the behavior type; for example, at time $t$, user $u$ added item $v_x$ to the cart, then at time $t+1$, the user marked item $v_y$ as a favorite. We regard the item behavior to provide contextual information, which indicates the user preferences on the platform. Thus we encode the sequence of contextual information and concatenate it to the item's latent features. To obtain the item embedding $v'_{j} \in \mathbb{R}^{d}$. Firstly the item one-hot-encoded vector $v_{j}$ is fed into a fully connected layer to generate the item embedding; 
\begin{equation}
    v'_{j}= v_j W_l +b_l
\end{equation} 
where $W_l \in \mathbb{R}^{I \times d}$ is the weight matrix, $I$ is the number of items, and $d$ is the items embedding dimension.
Secondly, the context (behavior types) one-hot-encoded vectors corresponding to the sequence are fed into a fully connected layer to obtain the context embedding;

\begin{equation}
    c'_{j}= c_j W_c +b_c
\end{equation}
where $W_c \in \mathbb{R}^{K \times d}$ is the weight matrix, $K$ is the number of behaviors, and $d$ is the context embedding dimension for simplicity, we use the same dimension for $d$ and $d$ and $b_c$ is the bias term. 

Accordingly, we can get the joint embedding by column-wise concatenating each item embedding with it's context embedding and feeding it to a fully connected layer;
\begin{equation}
    r_{j}= \text{concat}_{col} \left( v'_j, c'_j  \right)
\end{equation}
\begin{equation}
    q_{j}= r_j W_f +b_f
\end{equation}
where $W_f \in \mathbb{R}^{2d \times d}$ is the weight matrix, $d$ is the layer embedding dimension, and $b_f$ is the bias term. 
Finally, a learnable positional embedding $P_j$ is added to the item embedding to indicate the item's position within the historical sequence of interactions;
\begin{equation}
    e_{j}= q_{j} +P_j
\end{equation}

\subsection{Multi-head Self-Attention}
 Sequence-to-sequence modeling is essential in different fields, e.g., natural language processing and recommender systems, where the Transformer model \cite{vaswani2017attention} was firstly introduced for machine translation. One of the critical components of the transformer architecture is the multi-head self-attention block, which enables the model to learn the dependencies between the different sequence tokens, words, or items by applying a scaled-dot product. Here, we employ Multi-Head Self-Attention to encode each item in the user's sequence using all the other items in the input sequence. Enabling the encoded sequence to have rich and expressive representation. 
 
In this case, we concatenate the item sequence embedding row-wise $[e_1, e_2, ..., e_{|S^{u}_{t}|}]$ as shown in Figure \ref{fig:1} to form the input matrix $E \in \mathbb{R}^{  |{S^{u}_{t}}| \times d } $ for the multi-head attention block. Subsequently, we split the input into Query $(Q)$, Key $(K)$, and Value $(V)$ among the specified number of heads to apply the multi-head attention layer as follows: 
\begin{equation} 
\textrm{Att}(\textbf{Q}, \textbf{K}, \textbf{V}) = \textrm{softmax}\left( \frac{\textbf{Q} \textbf{K}^{T}}{\sqrt{\frac{d}{H}}}\right) \textbf{V}
\end{equation} 

\begin{equation} 
\begin{split}
&A=\textrm{SA}(E)= \\& 
\text{concat}_{col} \left( \textrm{Att}(E\textbf{W}^{Q}_{h}, E\textbf{W}^{K}_{h}, E\textbf{W}^{V}_{h}) \right)_{h=1:H}
\end{split}
 \label{eq:5}
\end{equation} 
where $\textbf{W}^{Q}_{h}$, $\textbf{W}^{K}_{h}$, $\textbf{W}^{V}_{h} \in \mathbb{R}^{d \times \frac{d}{H}}$ represent the linear projection matrices of the head at index $h$, and $H$ is the number of heads.
$A$ represents the column-wise concatenation of the attention heads.
Finally, we have the point-wise feed-forward layers to obtain the component's final output representations $Z \in \mathbb{R}^{  |{S^{u}_{t}}| \times d } $ as follows:
\begin{equation} 
\begin{split}
&Z = \textrm{FFN}(A) = \\&
\text{concat}_{row} \left( ReLU (A^{(n)} \textbf{W}^{(1)} + b^{(1)})\textbf{W}^{(2)} + b^{(2)}  \right)_{n=1:|{S^{u}_{t}}|}
\end{split}
 \label{eq:6}
\end{equation} 
\noindent where $\textbf{W}^{(1)}$, $\textbf{W}^{(2)} \in \mathbb{R}^{d \times d}$ are the weight matrices of the two feed-forward layers, and $b^{(1)}$, $b^{(2)} \in \mathbb{R}^{d}$ are their bias vectors. $\text{concat}_{row}$ concatenates vectors row-wise and $Z := [z_1, z_2, ..., z_{|S^{u}_{t}|}]$ is the latent output sequence of items after the multi-head self-attention block .

\subsection{Model Prediction}
After obtaining the item sequence interaction matrix after the multi-head self-attention block, we can stack several attention blocks to acquire higher-order interactions; however, in this work, one attention block received the best results. For scoring a target item at time $t+1$, we pass the target item id $v^o$ and the target behavior type $c^o$ through the same shared layers for item and context embedding. Then these embeddings are concatenated and passed through another shared embedding layer to obtain the final target item embedding $q^{o}_{|S^{u}_{t+1}|}$. Given the last item embedding in the interaction matrix $z_{|S^{u}_{t}|}$ and the target item embedding  $q^{o}_{|S^{u}_{t+1}|}$ the final score can be calculated as;

\begin{equation}
    \hat{Y} = \sigma(z_{|S^{u}_{t}|} \cdot q^{o}_{|S^{u}_{t+1}|})
\end{equation}
Where $\sigma$ is a sigmoid function.
\subsection{Model Learning}

To optimize our multi-behavior recommender model, we follow the same training process as SASRec
\cite{kang2018self} given its efficiency. We consider the input sequence as the original user interaction sequence excluding the last action $S^u=[ (v_1, c_1), (v_2, c_2), ..., (v_{ |S^{u}_{t-1}|},c_{|S^{u}_{t-1}|})]$. This sequence has a fixed maximum length that is attained by either padding with zeros\footnote{We assure that items IDs do not start from zero.} or truncating the sequence. Moreover, we build the target sequence as the right-shifted input sequence while including the last item interaction in the sequence. Similarly, the negative sampled input sequence is constructed by generating a sequence of random items which does not belong to the user's previous history $v \notin S^u$.
\subsubsection{Weighted binary cross entropy loss}
In the multi-behavior paradigm, there are different behaviors in the input sequence, but not all behaviors are equally crucial for the final model recommendations. Thus, we propose the weighted binary cross entropy loss in which we assign different weights for the different input behaviors. As shown in equation \ref{eq:9}, we propose adding the weighting factor $\alpha_b$ to our loss function, which can control the contribution of each behavior to the final loss. In this case, we have $K$ different $\alpha_b$s for each behavior in the given data, which means that the number of $\alpha_b$  can differ from one dataset to another. In the next section, an ablation study shows how the model performance varies with different weight settings. Furthermore, we use the weight $\beta$ to control the training process's positive/negative sampling ratio.
Hence our weighted binary cross entropy objective function for multi-behavior recommendation can be defined as follows;
\begin{equation}
      \mathcal{L}=  - \sum_{S^u \in \mathcal{S}} \sum_{t\in [1, 2, ..., |S^u|]} [ \alpha_b log{(\hat{Y}^{O^{(+)}}_{t})} + \beta (log{(1-\hat{Y}^{O^{(-)}}_{t})})]
       \label{eq:9}
\end{equation}
where $\hat{Y}^{O^{(+)}}_{t}$ are the output scores for the positive samples and $\hat{Y}^{O^{(-)}}_{t}$ are the output scores for the negative samples, $\mathcal{S}$ is the set of all sequences, $\alpha_b$ is the behaviors weights, and lastly, $\beta$ is the sampling weight. Finally, the model is optimized using ADAM optimizer \cite{kingma2014adam}. More details about the model hyper-parameters are described later.
\begin{table*}
\centering
\begin{tabular}{c|c|c|c|c}
\hline
Dataset     & Interactions & User\# & Item\# & Behavior \\
\hline
Taobao       & 7,658,926 & 147,894 & 99,037 & \textbf{Page View} (6,074,688), \textbf{Fav.} (267,131), \textbf{Cart} (674,191), \textbf{Buy} (642,916) \\
Yelp         & 1,400,000 & 19,800 & 22,734 & \textbf{Tip} (285,673), \textbf{Dislike} (198,106), \textbf{Neutral} (238,880), \textbf{Like} (6,773,43)  \\
MovieLens    & 9,922,036 & 677,88 & 8704 & \textbf{Dislike} (1,370,897), \textbf{Neutral} (3,580,155), \textbf{Like} (4,970,984) \\
Tianchi      &  4,619,389 & 25,000 & 500,900 & \textbf{Page View} (4,066,495), \textbf{Fav.} (251,984), \textbf{Cart} (4375), \textbf{Buy} (296,535)   \\ 
\hline
\end{tabular}
\caption{Datasets Statistics}
\label{tab:1}
\end{table*}

\begin{table*}[!]
\setlength{\tabcolsep}{4pt}
\centering
\begin{tabular}{c|cc|cc}
\hline

Method    & \multicolumn{2}{c}{Taobao} & \multicolumn{2}{c}{Yelp} \\
            & HR@10 & NDCG@10 & HR@10 & NDCG@10 \\
\hline  
\multicolumn{5}{c}{\textit{Sequential Recommendation Methods}}\\
\hline  
SASRec \cite{kang2018self} & 0.390 $\pm$ \tiny{$4.2E{-3}$} & 0.249 $\pm$ \tiny{$8E{-4}$}& 0.853  $\pm$ \tiny{$1.7E{-3}$}   &  0.5601 $\pm$ \tiny{$5.1E{-3}$}  \\

SSE-PT \cite{wu2020sse} & 0.393 $\pm$ \tiny{$2.7E{-3}$} & 0.232 $\pm$ \tiny{$3E{-4}$} & 0.857 $\pm$ \tiny{$1.5E{-2}$} &  0.572 $\pm$ \tiny{$1.8E{-2}$}  \\
\hline
\multicolumn{5}{c}{\textit{Context-Aware Recommendation Methods}}\\
\hline 
CARCA \cite{rashed2022context} & \underline{0.769 $\pm$ \tiny{$4E{-3}$}} &   \underline{0.662 $\pm$ \tiny{$5E{-3}$} } & 0.854 $\pm$ \tiny{$2E{-3}$}  & 0.589 $\pm$ \tiny{$4E{-3}$} \\
\hline
\multicolumn{5}{c}{\textit{Multi-Behavior Recommendation Methods}}\\
\hline 
MATN \cite{xia2020multiplex}& 0.354$\dagger$ & 0.209$\dagger$  & 0.826$\dagger$  & 0.530$\dagger$  \\
MB-GCN \cite{jin2020multi}& 0.369$\dagger$ & 0.222$\dagger$  &  0.796$\dagger$ & 0.502$\dagger$  \\
MB-GMN \cite{xia2021graph} & 0.491$\dagger$ & 0.300$\dagger$  & 0.861  $\pm$ \tiny{$8E{-3}$}  & 0.570  $\pm$ \tiny{$1.2E{-2}$} \\  
KHGT \cite{xia2021knowledge}& 0.464$\dagger$ & 0.278$\dagger$  & 0.880$\dagger$  & 0.603$\dagger$ \\
MB-STR \cite{yuan2022multi} & 0.768$\dagger$ & 0.608$\dagger$  & 0.882$\dagger$  & \underline{0.624}$\dagger$ \\
MBHT \cite{yang2022multi}& 0.745 $\pm$ \tiny{$6.1E{-3}$}& 0.559 $\pm$ \tiny{$8.4E{-3}$} & \underline{0.885 $\pm$ \tiny{$5E{-4}$}}  & 0.618 $\pm$ \tiny{$1.4E{-3}$} \\  
\hline
\textbf{CASM (ours)} & \textbf{ 0.841 $\pm$ \tiny{$3E{-4}$} } & \textbf{0.725 $\pm$ \tiny{$3E{-4}$}}  & \textbf{0.898 $\pm$ \tiny{$2E{-3}$}}  & \textbf{0.631 $\pm$ \tiny{$6E{-4}$}} \\
\hline  
Improv.(\%)& 9.50\% & 19.24\%  & 1.46\%  & 1.12\% \\
\hline 
\end{tabular}
\caption{Model performance and comparison against baselines on Taobao and Yelp datasets. Bold represents the best performance, and underline represents the second-best obtained results. Published results are indicated by the $\dagger$ symbol.}
\label{tab:2}
\end{table*}


\begin{table*}[!]
\setlength{\tabcolsep}{4pt}
\centering
\begin{tabular}{c|cc|cc}
\hline

Method    & \multicolumn{2}{c}{MovieLens}  & \multicolumn{2}{c}{Tianchi}\\
            & HR@10 & NDCG@10 & HR@10 & NDCG@10\\
\hline  
\multicolumn{5}{c}{\textit{Sequential Recommendation Methods}}\\
\hline  
SASRec \cite{kang2018self} &  0.911 $\pm$ \tiny{$1E{-3}$}& 0.668 $\pm$ \tiny{$5.1E{-3}$}& 0.659 $\pm$ \tiny{$3E{-3}$}& 0.495 $\pm$ \tiny{$2E{-3}$}  \\

SSE-PT \cite{wu2020sse}  & 0.911 $\pm$ \tiny{$7.1E{-3}$} & 0.657 $\pm$ \tiny{$4.5E{-3}$} & 0.663 $\pm$ \tiny{$1.2E{-2}$} & 0.468 $\pm$ \tiny{$1.3E{-2}$}  \\
\hline
\multicolumn{5}{c}{\textit{Context-Aware Recommendation Methods}}\\
\hline 
CARCA \cite{rashed2022context} & 0.906 $\pm$ \tiny{$2E{-3}$} & 0.665 $\pm$ \tiny{$1E{-3}$} &   0.713 $\pm$ \tiny{$4E{-4}$}&  0.500 $\pm$ \tiny{$1E{-3}$}\\
\hline
\multicolumn{5}{c}{\textit{Multi-Behavior Recommendation Methods}}\\
\hline 
MATN \cite{xia2020multiplex}& 0.847$\dagger$ & 0.569$\dagger$ & 0.714 $\pm$ \tiny{$7E{-4}$}  & 0.485 $\pm$ \tiny{$2E{-3}$} \\
MB-GCN \cite{jin2020multi}&  0.826$\dagger$ & 0.553$\dagger$ & -  & - \\
MB-GMN \cite{xia2021graph} &   0.820  $\pm$ \tiny{$1.1E{-3}$}& 0.530  $\pm$ \tiny{$9E{-4}$} & \underline{0.737  $\pm$ \tiny{$4.3E{-3}$}}  & 0.502  $\pm$ \tiny{$1.9E{-3}$} \\  
KHGT \cite{xia2021knowledge}& 0.861$\dagger$ & 0.597$\dagger$ & 0.652 $\pm$ \tiny{$1E{-4}$}  & 0.443 $\pm$ \tiny{$1E{-4}$} \\
MB-STR \cite{yuan2022multi}  & - & - & -  & - \\
MBHT \cite{yang2022multi} &  \underline{0.913 $\pm$ \tiny{$5.9E{-3}$}} & \underline{0.695 $\pm$ \tiny{$7E{-3}$}} & 0.725 $\pm$ \tiny{$6.3E{-3}$}  & \underline{0.554 $\pm$ \tiny{$4.8E{-3}$}} \\  
\hline
\textbf{CASM (ours)} &  \textbf{0.930 $\pm$ \tiny{$6E{-4}$}} & \textbf{0.713 $\pm$ \tiny{$1.3E{-3}$}} & \textbf{0.755 $\pm$ \tiny{$9E{-4}$}}  & \textbf{0.584 $\pm$ \tiny{$2.7E{-3}$}} \\
\hline  
Improv.(\%)& 1.86\% & 2.44\% & 2.99\%  & 5.95\% \\
\hline 
\end{tabular}
\caption{Model performance and comparison against baselines on MovieLens and Tianchi datasets. Bold represents the best performance, and underline represents the second-best obtained results. Published results are indicated by the $\dagger$ symbol.}
\label{tab:3}
\end{table*}


\section{Experiments}
In this section, we aim to answer the following research questions;\\
\begin{itemize}
\item \textbf{RQ1:} How does CASM model perform against the state-of-the-art multi-behavioral and sequential models?
\item \textbf{RQ2:} How do the contextual features affect the model performance?
\item \textbf{RQ3:} What is the effect of selecting different $\alpha_b$ for the different behaviors?
\item \textbf{RQ4:} What is the effect of using auxiliary behaviors on users with a limited number of interactions?

\end{itemize}
\subsection{Experimental Settings}
\subsubsection{Datasets}
We evaluate our model performance on four multi-behavioral datasets.\\
\textbf{Taobao \footnote{https://github.com/akaxlh/MB-GMN/tree/main/Datasets/Tmall} \cite{xia2021knowledge,yuan2022multi} :} is one of the most popular online shopping platforms in china. The datasets include four main behaviors of the platform users; buy, add-to-cart, add-to-favorite, and pageview. Where we consider the buy behavior as the target behavior.\\
\textbf{Tianchi \footnote{https://tianchi.aliyun.com/competition/entrance/231576/information} :} This data was initially provided for the Repeat Buyers Prediction Challenge, which is collected from Tmall.com. The dataset includes four behaviors similar to the Taobao data; buy, add-to-cart, add-to-favorite, and pageview.\\
\textbf{Yelp \footnote{https://github.com/akaxlh/KHGT/tree/master/Datasets/Yelp} \cite{xia2021knowledge,yuan2022multi} :} is gathered from the Yelp challenge in which the different behaviors are decided based on the users' explicit rating, which can vary between 1 and 5. As 1 is the worst rating and 5 is the best rating. Accordingly, the rating range is split into three ranges to obtain the three different behaviors; dislike (rating $\leq$ 2), natural (2< rating < 4), and like (rating $\geq$ 4). Additionally, the user can write a tip on the visited venue.\\
\textbf{MovieLens \footnote{https://github.com/akaxlh/KHGT/tree/master/Datasets/MultiInt-ML10M} \cite{xia2021knowledge} :} This dataset includes users' movie ratings that was handled similarly to the Yelp data with three behaviors dislike, neutral, and like. Whereas like behavior is considered as the main behavior for both datasets. Datasets statistics are summarized in Table \ref{tab:1}.

\subsection{Evaluation Protocol}
To guarantee a fair comparison, we follow the same evaluation protocol as current state-of-the-art methods \cite{xia2021knowledge, yuan2022multi}. We consider the users chronologically ordered sequence; we employ the leave-one-out mechanism, as we consider the whole sequence excluding the last interaction for training and validation, and preserve the last item for testing. For evaluation, we sample 99 negative items for each positive item. We utilize two widely used evaluation metrics, Hit Ration (HR@N) and Normalized Discounted Cumulative Gain (NDCG@N), where the higher the HR and NDCG, the more profitable the models' performances. To ensure the significance of the results, we report the mean and standard deviation of three different runs.

\subsubsection{Baselines}
We compare the proposed method against various recommendation methods to show our model superiority.\\

\paragraph{\textbf{Sequential Recommendation Methods.}}
\begin{itemize}
\item \textbf{SASRec} \cite{kang2018self}: A model that utilizes multi-head self-attention to capture the sequential pattern in the users' history, then applies dot product for calculating the items scores.  
\item \textbf{SSE-PT} \cite{wu2020sse}: A state-of-the-art model which incorporates the user embedding into a personalized transformer with stochastic shared embedding regularization for handling extremely long sequences.  
\end{itemize}
\paragraph {\textbf{Context-Aware Recommendation Methods.}}
\begin{itemize}
\item \textbf{CARCA} \cite{rashed2022context}: A state-of-the-art context-aware sequential recommendation approach that applies cross-attention between user profiles and items to predict the items' scores. To have a fair comparison against CARCA, we utilize the same contextual features of CASM as input.
\end{itemize}
\paragraph {\textbf{Multi-Behavior Recommendation Models.}}
\begin{itemize}
\item \textbf{MATN} \cite{xia2020multiplex}: A memory-augmented transformer network that utilizes a transformer-based multi-behavior encoder which jointly models behavioral dependencies.
\item \textbf{MB-GCN} \cite{jin2020multi}: This model represents the multi-behavioral data as a unified graph, then it employs a graph-convolutional network to learn the node's representation.  
\item \textbf{MB-GMN} \cite{xia2021graph}: Learns the multi-behavior dependencies through graph meta-network.
\item \textbf{KHGT} \cite{xia2021knowledge}: This model employs a knowledge graph hierarchical transformer to capture the type-wise behavior dependencies in the recommender model.
\item \textbf{MB-STR} \cite{yuan2022multi}: This model employs multi-behavior sequential transformer layers to learn sequential patterns across several behaviors. 
\item \textbf{MBHT} \cite{yang2022multi}: A hypergraph-enhanced transformer model which utilizes low-rank self-attention to model short-long behavioral dependencies. 
\end{itemize}


\begin{table*}
  \begin{minipage}{.5\linewidth}
    \centering
    \begin{tabular}{c|c|c|c|c|c }
      \toprule
       Buy & Cart & Fav & PV & NDCG@10 & HR10 \\
      \midrule
       1.0 & 0.0 & 0.0 & 0.0 & 0.5742  & 0.6779  \\
       0.9 & 0.1 & 0.0 & 0.0 & 0.6085  & 0.7199  \\
       0.8 & 0.1 & 0.1 & 0.0 & 0.6216	& 0.7407  \\
       \rowcolor{wildblueyonder}
       0.7 & 0.1 & 0.1 & 0.1 & \textbf{0.7254}	& \textbf{0.8411} \\
       0.6 & 0.2 & 0.1 & 0.1 &  0.7226	& 0.8399 \\
       0.5 & 0.2 & 0.2 & 0.1 &  0.7138	& 0.8386  \\
       0.4 & 0.3 & 0.2 & 0.1 &  0.7213 & 0.8390  \\
       0.3 & 0.3 & 0.3 & 0.1 &  0.7137	& 0.8363  \\
       0.3 & 0.3 & 0.2 & 0.2 &  0.7120	& 0.8360 \\
      \bottomrule
    \end{tabular}
    \caption{Different behaviors effect on Taobao Dataset}\label{tab:4}
  \end{minipage}%
  \begin{minipage}{.5\linewidth}
    \centering
   \begin{tabular}{c|c|c|c|c|c }
      \toprule
       Buy & Cart & Fav & PV & NDCG@10 & HR10 \\
      \midrule
       1.0 & 0.0 & 0.0 & 0.0 & 0.4152   &  0.6207\\
       0.9 & 0.1 & 0.0 & 0.0 & 0.4166	  & 0.6235 \\
       0.8 & 0.1 & 0.1 & 0.0 &  0.4280 & 0.6313 \\
       \rowcolor{wildblueyonder}
       0.7 & 0.1 & 0.1 & 0.1 &  \textbf{0.5873}	 & \textbf{0.7561} \\
       0.6 & 0.2 & 0.1 & 0.1 &  0.5867	 & 0.7537 \\
       0.5 & 0.2 & 0.2 & 0.1 & 0.5840  & 0.7517 \\
       0.4 & 0.3 & 0.2 & 0.1 & 0.5844 	 & 0.7579 \\
       0.3 & 0.3 & 0.3 & 0.1 & 0.5869  & 0.7552 \\
       0.3 & 0.3 & 0.2 & 0.2 &  0.5813	 & 0.7510 \\
      
      \bottomrule
    \end{tabular}
    \caption{Different behaviors effect on Tianchi Dataset}\label{tab:5}
    			
  \end{minipage}
\end{table*}

\begin{table*}
  \begin{minipage}{.5\linewidth}
    \centering
    \begin{tabular}{c|c|c|c|c|c }
      \toprule
       Like & Tip & Neutral & Dislike & NDCG@10 & HR10 \\
      \midrule
       1.0 & 0.0 & 0.0 & 0.0 & 0.5935	  &  0.8709  \\
       0.9 & 0.1 & 0.0 & 0.0 &  0.6045	 & 0.8857   \\
       0.8 & 0.1 & 0.1 & 0.0 & 0.6170	  &  0.8902  \\
       0.7 & 0.1 & 0.1 & 0.1 &  0.6250	 &  0.8957  \\
       0.6 & 0.2 & 0.1 & 0.1 & 0.6247  & 0.8973 \\
       0.5 & 0.2 & 0.2 & 0.1 &  0.6240	 & 0.8966   \\
       0.4 & 0.3 & 0.2 & 0.1 & 0.6249	  &  0.8965  \\
       0.3 & 0.3 & 0.3 & 0.1 & 0.6237	  &  0.8954  \\
       \rowcolor{wildblueyonder}
       0.3 & 0.3 & 0.2 & 0.2 &  \textbf{0.6276}	 & \textbf{0.8978}   \\
      
      \bottomrule
    \end{tabular}
    \caption{Different behaviors effect on Yelp Dataset}\label{tab:6}
  \end{minipage}%
  \begin{minipage}{.5\linewidth}
    \centering
   \begin{tabular}{c|c|c|c|c }
      \toprule
       Like & Neutral & Dislike  & NDCG@10 & HR10 \\
      \midrule
       1.0 & 0.0 & 0.0 & 0.6972  & 0.9219  \\
       \rowcolor{wildblueyonder}
       0.9 & 0.1 & 0.0 & \textbf{0.7124} & \textbf{0.9300}  \\ 	
       0.8 & 0.1 & 0.1 & 0.6971 & 0.9210  \\
       0.7 & 0.1 & 0.1 & 0.6914  & 0.9222  \\
       0.6 & 0.2 & 0.1 & 0.6878  & 0.9198  \\
       0.5 & 0.2 & 0.2 & 0.6943  & 0.9200  \\
       0.4 & 0.3 & 0.2 & 0.6905  & 0.9195  \\
       0.3 & 0.3 & 0.3 & 0.6878  & 0.9174  \\
       0.3 & 0.3 & 0.2 & 0.6879  & 0.9163 \\
      	
      \bottomrule
    \end{tabular}
    \caption{Different behaviors effect on MovieLens Dataset}\label{tab:7}
  \end{minipage}
\end{table*}

\subsection{Model Performance (\textbf{RQ1})}
 We conducted experiments to evaluate our proposed model CASM against several state-of-the-art approaches. As discussed earlier, we compare our model against various recommendation models, not only multi-behavioral recommendation methods but also methods used for sequential recommendation and a context-aware model, as they are highly relevant approaches. 
We analyze the model performance against the sequential recommendation methods; we selected two state-of-the-art sequential models to compare their performance against multi-behavior methods, reflecting the importance of utilizing the other behaviors as they employ only single behavior. Looking into SASRec and SSE-PT results, they show very high competitive results against the state-of-the-art multi-behavioral models. On the other hand, methods such as MATN, MB-GCN, MB-GMN, and KHGT show fragile results, specifically on the Taobao dataset, and other models, essentially SASRec, MB-STR, MBHT, which emphasize the significance of modeling the sequential patterns in the input data. Nevertheless, our model CASM significantly outperforms all types of baselines and improves the HR on the Taobao dataset by 9.5\% and 2.99\% on the Tianchi dataset. Additionally, context-aware models, which incorporate the time aspect, can also be a strong competitor as they benefit from both time and context in their sequential approach. In this case, we compete against the state-of-the-art context-aware method CARCA and our model CASM shows significantly better results on the four data sets with a reasonable margin. Regardless, the CARCA model shows competitive results as it benefits from both sequential and similar contextual information that exist in the data.

Moreover, we compare the CASM model to the current state-of-the-art multi-behavioral models. Graph-based methods successfully learn the user-item and item-item dependencies. Nevertheless, they fail to capture the temporal dynamics in the data. Consequently, the results for MB-GCN, MB-GMN, and KHGT show highly competitive results on Yelp and MovieLens datasets where the user-item and item-item dependencies play essential roles. Nevertheless, they fail to capture the information in the Taobao dataset. Recent approaches such as MB-STR and MBHT employ multi-head self-attention mechanisms that allow the model to capture the sequential pattern in the data and learn item-item dependencies. However, our proposed model CASM shows superior performance against these recent multi-behavioral methods with significantly less runtime as the model benefits from the multi-head self-attention layers in capturing sequential patterns. Furthermore, applying the weighted binary cross entropy loss allows the model to control the contribution of different behaviors according to their actual contribution to the next item recommendation task.  

\begin{figure}
\centering
\begin{minipage}{.42\textwidth}
  \centering
   \includegraphics[scale=0.25]{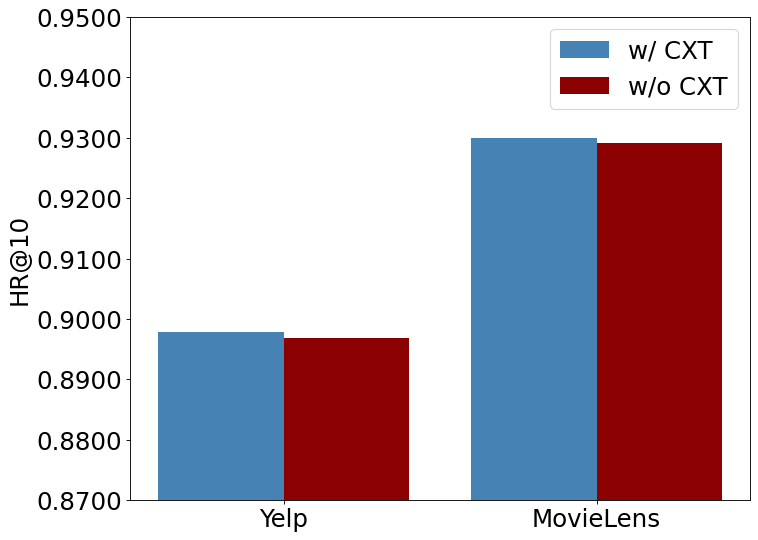}     
  \label{fig:2}
\end{minipage}%
\begin{minipage}{.42\textwidth}
  \centering
   \includegraphics[scale=0.25]{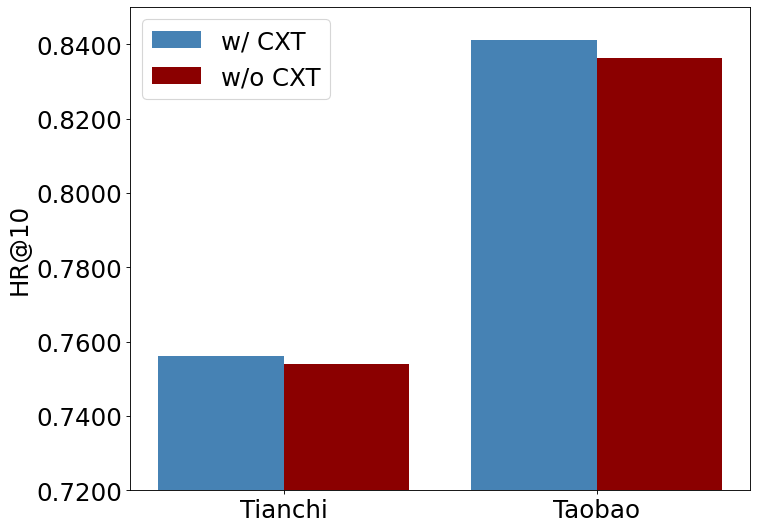}         
  \label{fig:3}
\end{minipage}

\caption{Effect of adding contextual features on Tianchi/Taobao and Yelp/MovieLens datasets HitRatio and NDCG}
\end{figure}

\subsection{Ablation Study: Effect of Contextual Features (\textbf{RQ2})}
We investigate the effectiveness of different components added to the proposed model. Initially, we study the influence of adding contextual features. In this case, we removed the contextual features (behaviors types) concatenated to the items embedding to form the final item latent features. As shown in Figure 2, the effect of behavior types on the HR of the Yelp and MovieLens dataset is minimal. This can be explained by the nature of the behaviors in the data as these behaviors datasets are initially created based on ratings where the behaviors are hand-crafted, which is why adding behavior types is ineffective. On the other hand, the effect of including behavior types can be seen in Tianchi and Taobao datasets. The HR slightly increased after incorporating the behavior types. However, the impact is smaller compared to other components investigated in the subsequent sections.

\begin{figure*}
\centering
\begin{minipage}{.51\textwidth}
  \centering
   \includegraphics[scale=0.125]{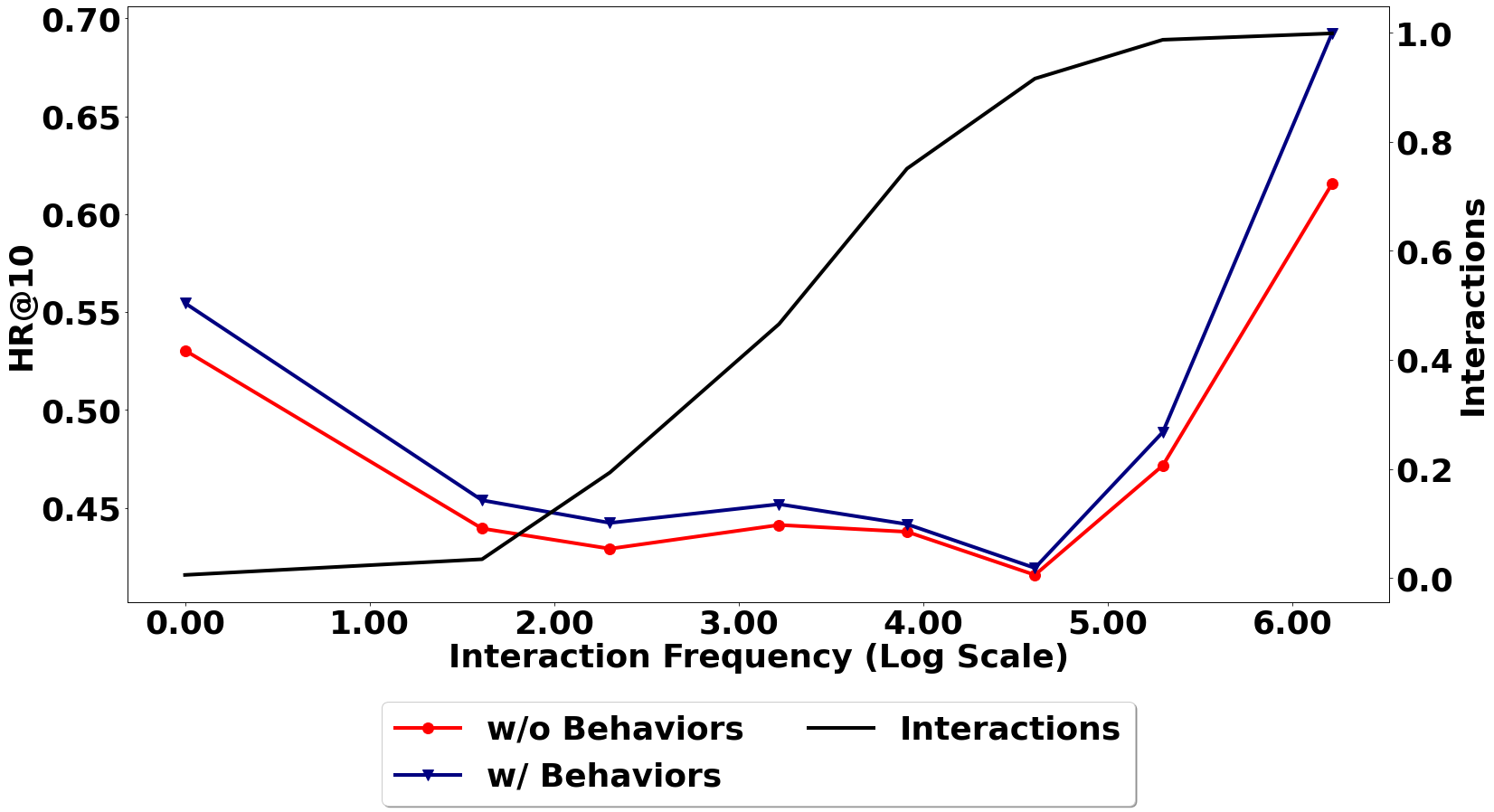}     
\end{minipage}%
\begin{minipage}{.51\textwidth}
  \centering
   \includegraphics[scale=0.125]{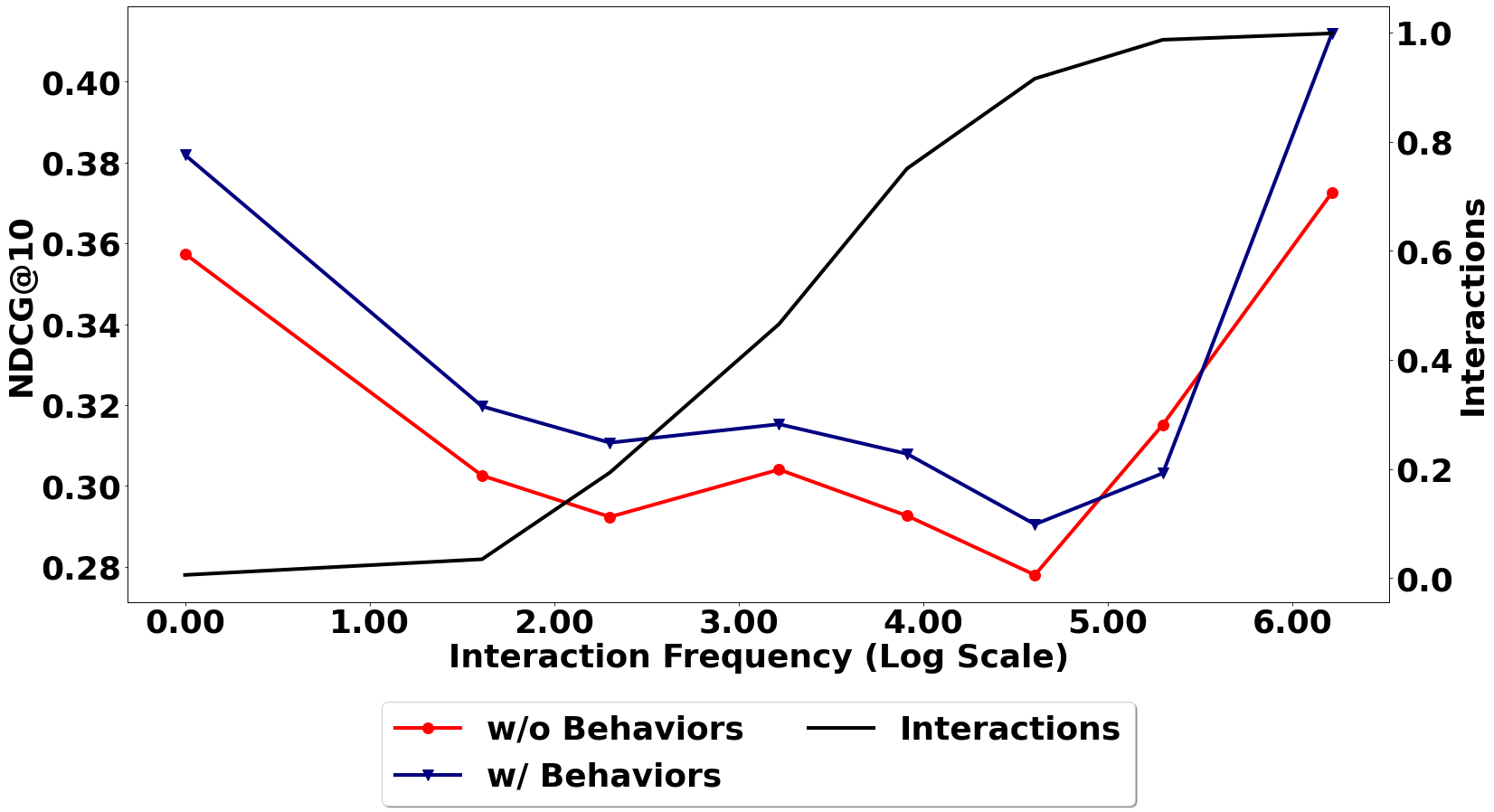}       
  \label{fig:3}
\end{minipage}

\caption{Effect of adding auxiliary behaviors of different users on Yelp dataset HitRatio and NDCG}

\begin{minipage}{.51\textwidth}
  \centering
   \includegraphics[scale=0.125]{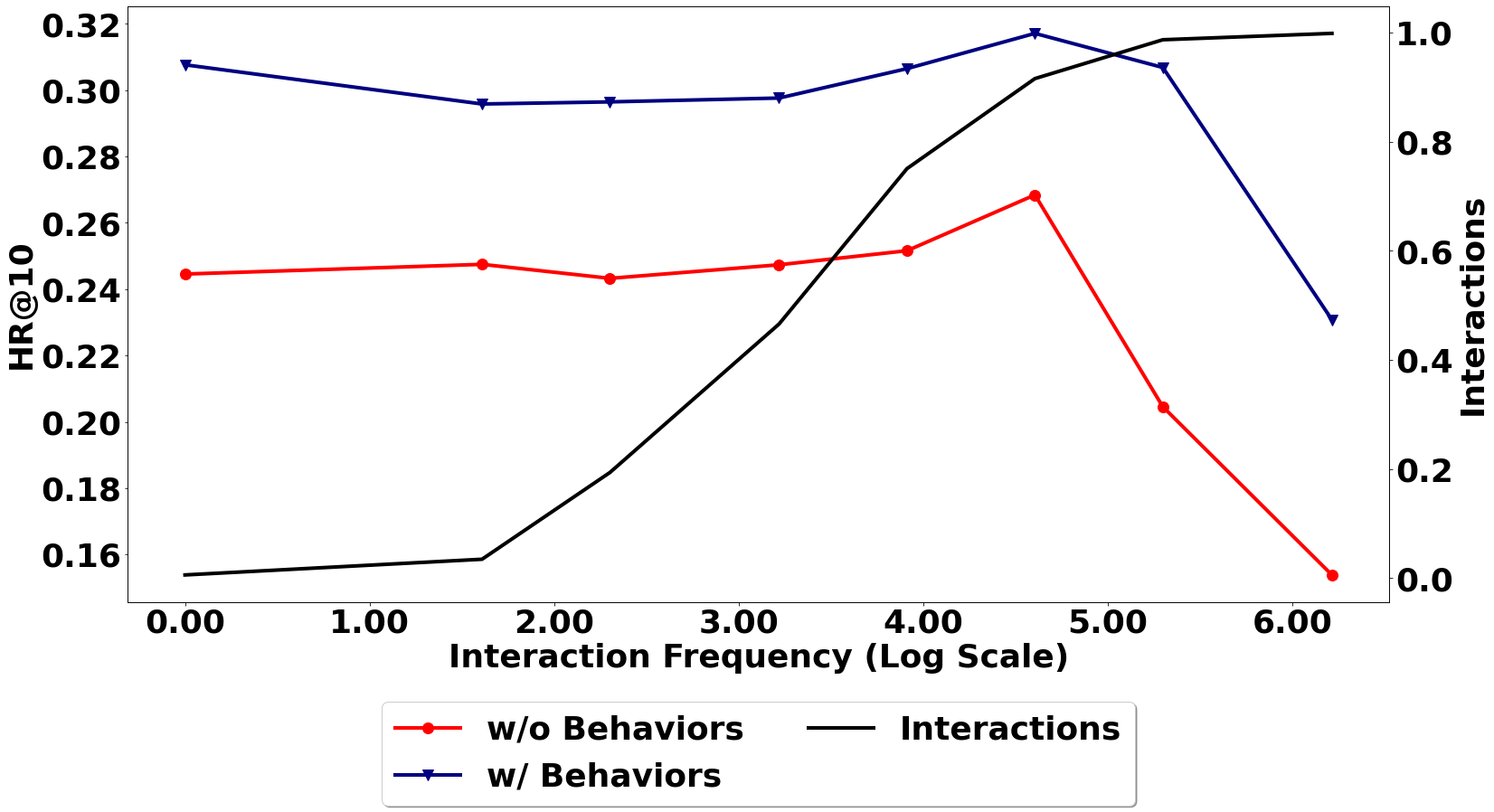}     
\end{minipage}%
\begin{minipage}{.51\textwidth}
  \centering
   \includegraphics[scale=0.125]{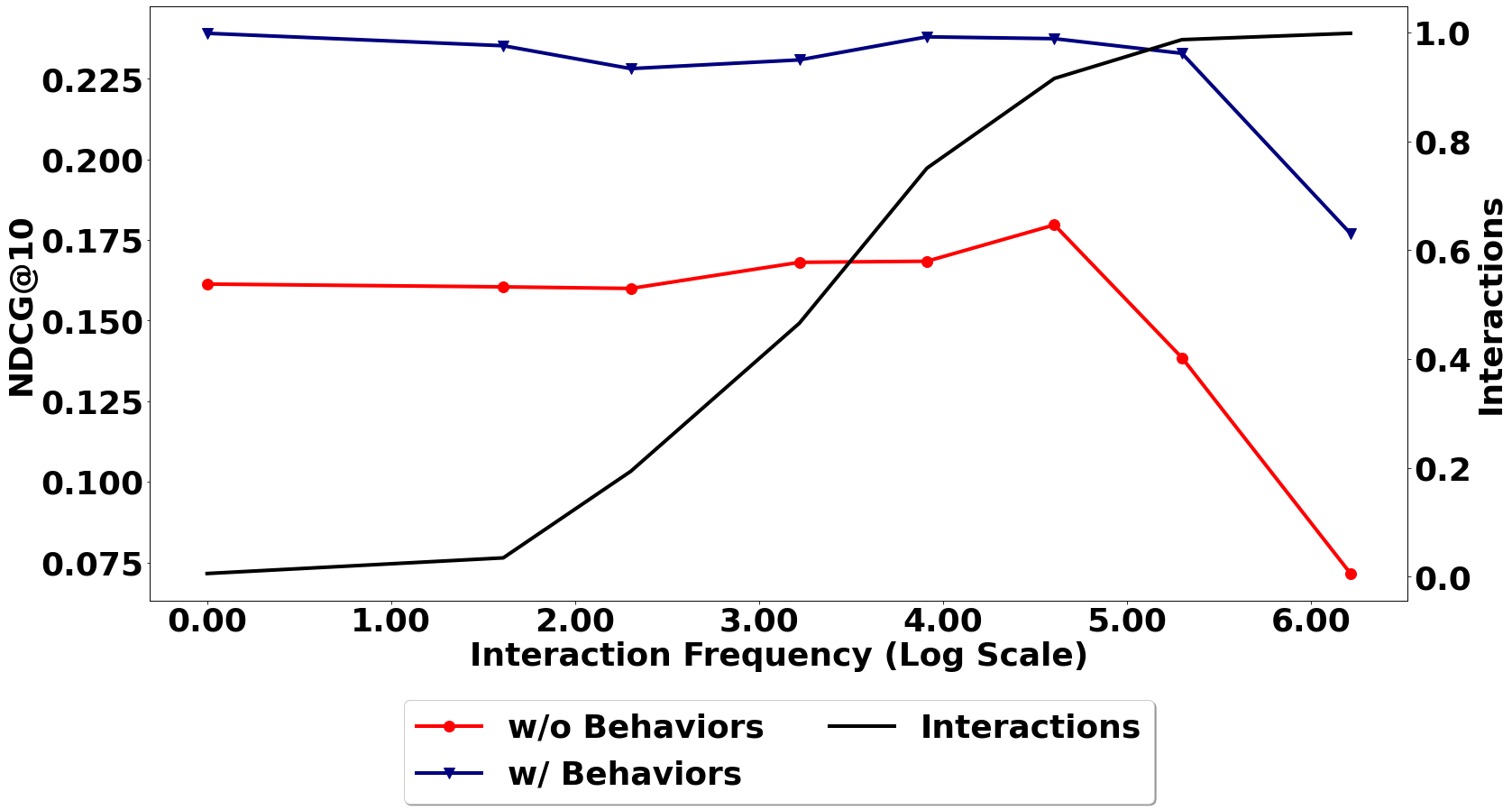}     
 \label{fig:4}
\end{minipage}
 
 \caption{Effect of adding auxiliary behaviors of different users on Tianchi dataset HitRatio and NDCG}
\end{figure*}

\subsection{Effect of Selecting Weighting Factor $\alpha_b$ (\textbf{RQ3})}

As discussed earlier, multi-behavioral data includes various user-item interaction relations, meaning that the user can interact with the same item or different items in several ways. Naturally, the main target of the recommendation task is to give higher scores to the items the user will most likely interact with. For instance, if we consider all the user-item interactions from different behaviors equally while focusing on the primary behavior as our main recommendation objective, we implicitly model all behaviors with equal significance. Nevertheless, this is not always the case with item recommendations; if the user views an item, it does not imply that it has the same importance as if he/she added it to the shopping cart. Here we utilized the factor $\alpha_b$ for each behavior in the datasets, we tuned the $\alpha_b$, which is not a full grid search, but we included the most critical weights that can affect the model performance. Table \ref{tab:7} shows the results on the MovieLens dataset, and it is clear that the best performance is obtained when setting the weights to 0.9, 0.1, and 0 for the like, neutral, and dislike behaviors, respectively. This clarifies that the behaviors for this data do not play a critical role like in other datasets such as Tianchi dataset.
On the contrary, Tianchi and Taobao datasets best weight settings are 0.7, 0.1, 0.1, and 0.1 for the buy, cart, favorite, and page views, respectively. Therefore, incorporating the behaviors' weights significantly affects multiple datasets. In Tianchi, the hit ratio bounced from 0.6207 to 0.7561 and from 0.6779 to 0.8411 on the Taobao dataset, which reflects the significance of including the page view behavior in both datasets despite the low weight set to the cart, favorite, and page view behaviors. Nevertheless, we tried to learn these behaviors with larger weights, which did not affect the results positively. Finally, for the Yelp dataset, tuning the weights had a 2\% performance improvement on the HR and a 3\% improvement on the NDCG, which is also a considerable improvement. To conclude, considering auxiliary behaviors in the model can significantly affect the model performance, especially if the behaviors are expressive. However, the behaviors can have minimal effect if the behaviors are not correctly defined.

\subsection{Effect of Adding Auxiliary Behaviors on Users with Limited Number of Interactions (\textbf{RQ4})}
We further analyze the effect of adding auxiliary behaviors on the user level. First, we sort the users based on the frequency of their primary interaction. Then we analyze the model performance before and after incorporating auxiliary behaviors. For the Yelp dataset, the interactions represent the count of the \textit{like} behavior interactions. As shown in Figure 3, adding behaviors consistently show better HR results for various users.
Similarly, the NDCG has shown an even higher improvement margin except for a few users, which tend to have a higher number of interactions. On the other hand, Figure 4 displays the effect of adding the auxiliary behaviors on different users in the Tianchi dataset; from the figure, we can observe the massive improvement in the HR and NDCG consistently for all user types for users with a low number of \textit{buy} interaction increased from 0.245 to 0.31. In contrast, the NDCG increased from 0.160 to 0.23, which justifies the importance of auxiliary behaviors on the user-level scope.

\begin{table}[H]
\centering
\begin{tabular}{c|c}
\hline
Model     & Average batch runtime (sec)  \\
\hline
KHGT      &  7.830  \\
MBHT      &  0.957  \\
MB-GMN    &  0.259  \\
CARCA     &  0.056  \\
\hline
CASM (ours)      &  \textbf{0.029}  \\
\hline
\end{tabular}
\caption{Runtime analysis on Yelp dataset}
\label{tab:8}
\end{table}

\section{Hyperparameters Settings}
We ran our experiments using GPU RTX 2070 Super and CPU Xeon Gold 6230 with RAM 256 GB. We used Tensorflow\footnote{https://www.tensorflow.org} for implementation;  For hyperparameter optimization, we adopted a grid search on the latent embedding size between [30 - 120], on the learning rate between [0.001 and 0.00005], the maximum sequence length between [50 and 200], the batch size we tested [128 and 256], and the dropout rate between [0.2 and 0.6]. The best parameters for dropout are 0.25, a maximum length of 150, an embedding size of 85, and a learning rate of 0.0005, respectively, for the Taobao dataset, while for the Tianchi dataset, the dropout is 0.5, the maximum length is 70, embedding size of 50, and learning rate 0.0007. MovieLens dataset dropout is 0.4, the maximum length is 70, the embedding size is 70, and the learning rate is 0.0006. Finally, for the Yelp dataset,  the dropout is 0.5, the maximum length is 150, the embedding size is 50, and the learning rate is 0.0003. For all datasets, we employed one attention block with one head with a batch size of 128, as it showed the best performances in our case. Furthermore, we used the ReLU activation function for all the models' layers and the ReLU activation function in the feed-forward layer in the multi-head self-attention block, which has shown better performance. We tuned the $\alpha_b$ between [0 and 1]; the next section demonstrates an ablation study of the best settings. Furthermore, we tuned the $\beta$ factor between [0 and 2], where 1.1 showed the best results. Finally, we employ Adam\cite{kingma2014adam} for optimizing the proposed model. Our implementation code is available  \footnote{\url{https://github.com/Shereen-Elsayed/CASM}}.

\section{Runtime and Scalability Analysis}
Runtime and model complexity are crucial aspects that many models try to tackle. We analyze the runtime of our proposed model CASM versus the current state-of-the-art models. For a fair comparison, we fix the batch size for all models to size 128 on the Yelp dataset to calculate the time consumed by the model for single batch processing, i.e., we report the average of three batches. As shown in Table \ref{tab:8} CASM model has 99.7 \% improvement over the KHGT model, 97.7 \% over MBHT, and 60.7\% improvement over the second best model CARCA. This proves the efficiency of our model runtime and simplicity versus all state-of-art-models. 

We add further ablation study about the model scalability to show how the model scales given different input sequence lengths. Table \ref{tab:9} illustrates the model batch processing time and performance given various sequence lengths on the Yelp dataset. The model performance starts to saturate starting sequence length of 150 with HR 0.898 and NDCG 0.631. Furthermore, the model time scales linearly for the different sequence lengths. However, our model given larger sequence lengths, shows more than 35 times faster than the closest competitor MBHT \cite{yang2022multi}.

\begin{table*}[!]
\setlength{\tabcolsep}{4pt}
\centering
\begin{tabular}{ccccccccc}
\hline

$|{S^{u}_{t}}|$    & 20 & 50 & 70 & 100 & 150 & 200 & 300 & 400  \\
\hline  
Time (sec )  & 0.015 & 0.017 & 0.022 & 0.024 & 0.029 & 0.039 & 0.064 &  0.096 \\
\hline  
HR@10   & 0.872 & 0.892 & 0.896 & 0.897 & 0.898 & 0.900 & 0.901 & 0.902 \\
NDCG@10 & 0.585 & 0.616 & 0.627 & 0.628 & 0.631 & 0.630 & 0.632 & 0.631  \\
\hline
\end{tabular}
\caption{Model performance on Yelp dataset given different sequence lengths.}
\label{tab:9}
\end{table*}

\section{Conclusion}
In this work, we presented a context-aware sequential model CASM that captures the sequential patterns in multi-behavioral data. First, the model encodes the input item sequence and benefits from incorporating the behavior types as contextual features that are concatenated with the item latent features. Then it employs a multi-head self-attention mechanism to capture item dependencies. Finally, we empower our model by proposing a weighted binary cross entropy loss that weights the behaviors differently based on their significance. Experimental results on four multi-behavioral datasets show that our model outperforms state-of-the-art recommendation methods.

\bibliographystyle{ACM-Reference-Format}
\bibliography{CASM-manuscript}

\appendix



\end{document}